\documentclass[9pt,twocolumn,twoside]{opticajnl}
\journal{opticajournal} 

\setboolean{shortarticle}{true}

\usepackage{lineno}
\usepackage{color}
\nolinenumbers 

\newcommand{\mtW}{\mathcal{W}}

\title{Wigner picture of partially coherent accelerating beams}
\author[1,2,3]{Sergey A. Ponomarenko}
\author[1]{Morteza Hajati}
\affil[1]{Department of Electrical and Computer Engineering, Dalhousie University, Halifax, Nova Scotia, B3J 2X4, Canada}
\affil[2]{Department of Physics and Atmospheric Science, Dalhousie University, Halifax, Nova Scotia, B3H 4R2, Canada}
\affil[3]{serpo@dal.ca}

\dates{Compiled \today}

\doi{\url{http://dx.doi.org/10.1364/OL.XX.XXXXXX}}

\begin{abstract}
We advance a phase-space theory of partially coherent accelerating, non-diffracting beams employing the Wigner distribution function (WDF). We derive a general expression for the WDF of any accelerating, diffraction-free beam of arbitrary degree of spatial coherence and find an elegant closed-form expression for the WDF of such beam with a Gaussian energy spectrum of noise. We also show how partially coherent accelerating beams of finite power can be constructed within the Wigner picture. 
\end{abstract}

\setboolean{displaycopyright}{false} 

\begin{document}

\maketitle

The momentous discovery of nondiffracting quantum wave packets that can accelerate in free space~\cite{balazs1979nonspreading} has prompted initial interest in the subject of wave packet acceleration. The subsequent experimental realization of finite-power, accelerating Airy beams, capable of defying diffraction over finite distances~\cite{siviloglou2007accelerating}, has triggered a flurry of research activity into accelerating optical beams~\cite{efremidis2019airy}. Coherent Airy beams have since found numerous applications to filamentation~\cite{polynkin2009curved}, supercontinuum generation~\cite{ament2011supercontinuum}, optical light bullet formation~\cite{chong2010airy},
optical imaging~\cite{jia2014isotropic,vettenburg2014light}, particle manipulation~\cite{grier2003revolution,baumgartl2008optically}, Airy plasmon generation~\cite{salandrino2010airy,minovich2011generation} and material processing~\cite{mathis2012micromachining}, among others~\cite{efremidis2019airy}. 

At the same time, much less work has been devoted to partially coherent diffraction-free beams~\cite{ponomarenko2007dark,lumer2015incoherent,hajati2021airy,cao2025diffraction}. In this context, Airy beams on incoherent background, which defy diffraction in free space at the expense of having their acceleration suppressed, were theoretically proposed~\cite{hajati2021airy} and experimentally realized~\cite{chen2024experimental}, and their self-healing properties were examined~\cite{chen2024experimental}. On the other hand, several particular classes of partially coherent Airy beams that maintain their acceleration in free space were introduced and even nearly incoherent accelerating Airy beams were generated in the laboratory~\cite{lumer2015incoherent}. 

In this Letter, we advance a general theory of accelerating beams of any degree of spatial coherence employing the concept of the Wigner distribution function (WDF). Specifically, we derive a general expression for the WDF of any accelerating, diffraction-free beam of arbitrary degree of spatial coherence and find an elegant closed-form expression for the WDF of such a beam with a Gaussian energy spectrum of noise. Our phase-space approach enables us to treat coherent and partially coherent accelerating beams within the same formalism. Our methodology brings new insight into the phase-space evolution of finite-power accelerating beams, and highlights an alternate route to the generation of such beams by truncating the momentum distributions of ideal accelerating beams. Our work may find applications to particle trapping and manipulation as well as to optical communications through noisy environments.

We start by recalling that the propagation of a paraxial wave packet of a classical optical field in free space is governed by the same Schr\"{o}dinger equation that describes the evolution of a quantum particle there. It follows that the evolution of a classical state vector $|\Psi\rangle$ in Hilbert space obeys the equation
\begin{equation}\label{SEQ}
    i\partial_z|\Psi\rangle=(\hat{k}^2/2)|\Psi\rangle,
\end{equation}
where $\langle x|\Psi\rangle$ is the electric field envelope of the wave packet and $\hat{k}=-id/dx$ in the coordinate representation. Henceforth, we work in dimensionless variables assuming any transverse spatial scale $x$ is normalized to a characteristic transverse width of a wave packet. A formal solution to Eq.~(\ref{SEQ}) reads
\begin{equation}
    |\Psi (z)\rangle=e^{-iz\hat{k}^2/2}|\Psi (0)\rangle,
\end{equation}
where, in the coordinate representation, we can express the field envelope as
\begin{equation}\label{Psi-x}
    \Psi(x,z)=\langle x|\Psi(z)\rangle=\langle x|e^{-iz\hat{k}^2/2}|\Psi(0)\rangle.
\end{equation}

As was shown elsewhere~\cite{unnikrishnan1996uniqueness}, the evolution operator for a 1D accelerating nonspreading wave packet must satisfy the equation
\begin{equation}\label{EvU}
    e^{-iz\hat{k}^2/2}=e^{iz^3/6}e^{izx}e^{-i\hat{k}z^2/2}e^{-iz\hat{H}(k,x)},
\end{equation}
where the Hamiltonian ought to have the form
\begin{equation}
    \hat{H}=\hat{k}^2/2+\hat{x}.
\end{equation}
Introducing the eigenfunctions of $\hat{H}$ corresponding to the eigenvalues $\varepsilon$,
\begin{equation}
    \hat{H}|\Psi_{\varepsilon}(0)\rangle=\varepsilon|\Psi_{\varepsilon}(0)\rangle,
\end{equation}
we show in the Supplement that the eigenfunctions in the coordinate representation are just scaled shifted Airy functions, such that
\begin{equation}\label{Ai}
    \Psi_{\varepsilon}(x,0)=Ai [2^{1/3} (x-\varepsilon)].
\end{equation}
Next, we can infer from Eqs.~(\ref{Psi-x}) and~(\ref{EvU}) that
\begin{equation}\label{Psi-acc}
    \Psi_{\varepsilon}(x,z)=e^{iz^3/6}e^{izx}e^{-i\varepsilon z}\Psi_{\varepsilon}(x-z^2/2,0).
\end{equation}
Eqs.~(\ref{Ai}) and (\ref{Psi-acc}) describes the only 1D accelerating non-spreading wave packet, either classical or quantum, originally discovered by Balasz and Berry~\cite{balazs1979nonspreading} in the context of quantum mechanics; the uniqueness was proven in~\cite{unnikrishnan1996uniqueness} with the aid of the operator approach that we adopt here. 

We can now consider partially coherent accelerating beams. To this end, we introduce a statistical operator $\hat{W}$~\cite{ponomarenko2002diffusion,ponomarenko2015self}, such that its matrix element yields a cross-spectral density of the classical light field, $\langle x_1|\hat{W}(z)|x_2\rangle=W(x_1,x_2,z)$. It stands to reason to represent the statistical operator in terms of coherent accelerating states, which are eigenstates of $\hat{H}$. To this end, we can express $\hat{W}$ by analogy with a quantum density operator as
\begin{equation}\label{hW}
    \hat{W}(z)=\int d\varepsilon\,p(\varepsilon) |\Psi_{\varepsilon} (z)\rangle\langle\Psi_{\varepsilon}(z)|.
\end{equation}
Here $p(\varepsilon)$ is a noise spectrum of a partially coherent wave packet, which we assume to be normalized, such that $\int d\varepsilon\,p(\varepsilon)=1$. Hereafter, all integrations are over the entire real axis. In the coordinate representation, Eq.~(\ref{hW}) reads
\begin{equation}
    W(x_1,x_2,z)=\int d\varepsilon\,p(\varepsilon)\Psi_{\varepsilon}^*(x_1,z)\Psi_{\varepsilon}(x_2,z),
\end{equation}
which is a familiar representation of the cross-spectral density in terms of pseudo-modes~\cite{gori2007devising,ponomarenko2011complex}, chosen to be continuously shifted Airy wave packets. We note in passing that a particular case of this representation with a discrete noise spectrum, $p(\varepsilon)=\sum_n p_n\delta(\varepsilon-\varepsilon_n)$, was employed in~\cite{lumer2015incoherent} to experimentally realize (nearly) incoherent accelerating beams. 

We now demonstrate that any accelerating partially coherent wave packet that resists diffraction can be elegantly described within the framework of the Wigner distribution function $\mtW$. The WDF formalism has been applied to treat quantum~\cite{leonhardt1997measuring} and classical~\cite{lohmann1993image,alonso2011wigner} wave packet evolution alike. Recently, the WDF has become instrumental in unveiling a new type of classical entanglement, phase-space nonseparability~\cite{ponomarenko2021twist}, which has been shown to dramatically affect optical beam shifts in reflection of structured wave packets from a flat interface~\cite{chen2025phase}. Defining the WDF as
\begin{equation}
    \mtW(X,k,z)=\int dx\,W(X-x/2,X+x/2,z)e^{-ikx},
\end{equation}
where we introduce the center-of-mass and difference coordinates
\begin{gather}
    X=(x_1+x_2)/2; \quad x=x_2-x_1,
\end{gather}
we can express, after some algebra, the Wigner function of such a wave packet as
\begin{equation}\label{Wig-gen}
    \mtW (X,k,z)=\int d\varepsilon\,p(\varepsilon)\mtW_{\mathrm{Ai}}(X-z^2/2-\varepsilon,k-z).
      \end{equation}
In Eq.~(\ref{Wig-gen}), the Wigner function of a coherent shifted Airy wave packet $\mtW_{\mathrm{Ai}}$, which we derive in the Supplement, reads
\begin{equation}\label{Wig-Ai}
    \mtW_{\mathrm{Ai}}(X-z^2/2-\varepsilon,k-z)=2^{-1/3}Ai(2X+k^2-2kz-2\varepsilon).
\end{equation}
Eqs.~(\ref{Wig-gen}) and~(\ref{Wig-Ai}) are among the key results of this work; they show that the WDF of an accelerating nonspreading wave packet of any degree of spatial coherence stems from a weighted superposition of WDFs of shifted replicas of a coherent Airy beam. 

However, Eq.~(\ref{Wig-gen}) describes an ideal accelerating partially coherent wave packet, which has a flat distribution in momentum space and hence carries infinite power. We now show how finite-power accelerating beams can be constructed employing the Wigner function. To this end, we consider a cut-off in momentum space of a source that engenders an ideal accelerating partially coherent beam. The cutoff function $f_a(k)$, where $a$ is a characteristic cut-off scale in k-space, must be integrable in momentum space, guaranteeing a finite total power of the source. The corresponding Wigner function of the source has the form
\begin{equation}\label{Wig-cut}
    \mtW_{\mathrm{FP}}(X,k,0)=f_a(k)\mtW(X,k,0),
\end{equation}
where the subscript FP stands for finite power. Next, the evolution of the Wigner function in free space is a pure phase-space shear, such that
\begin{equation}\label{Wig-ev}
    \mtW(X,k,z)=\mtW(X-kz,k,0).
\end{equation}
Combining Eqs.~(\ref{Wig-gen}), ~(\ref{Wig-Ai}), ~(\ref{Wig-cut}) and~(\ref{Wig-ev}), we can establish the following exceptionally simple evolution law for a finite-power accelerating wave packet:
\begin{equation}\label{Wig-fp}
 \mtW_{\mathrm{FP}}(X,k,z)=f_a (k)\mtW(X,k,z).
\end{equation}
Thus, constraining a source in momentum space ensures that any accelerating wave packet generated by the source can be realized in the laboratory. We note that the momentum cut-off can involve either fine or large scales in k-space. The former case is equivalent to introducing a (possibly soft) aperture in the source plane that was discussed before~\cite{siviloglou2007accelerating}, while the latter was precisely the protocol behind the first experimental realization of coherent finite-power Airy beams using an SLM~\cite{siviloglou2007observation}. We stress that Eq.~(\ref{Wig-fp}) is applicable to {\it any} accelerating paraxial wave packet that carries finite power.  

Further, it is no surprise that the momentum distribution of any accelerated, diffraction-free partially coherent beam in any transverse plane $z=const$ is entirely determined by the momentum-space cut-off, as evidenced by the corresponding marginal distribution of the Wigner function,
\begin{equation}
    J(k,z)=\int dX\,\mtW_{\mathrm{FP}}(X,k,z)\propto f_a(k),
\end{equation}
see the Supplement for details. The intensity, on the other hand, is given by the other marginal distribution:
\begin{equation}\label{I-def}
    I(X,z)=\int\frac{dk}{2\pi}\mtW_{\mathrm{FP}}(X,k,z).
\end{equation}

\begin{figure}[ht]
\centering
\includegraphics[width=\linewidth]{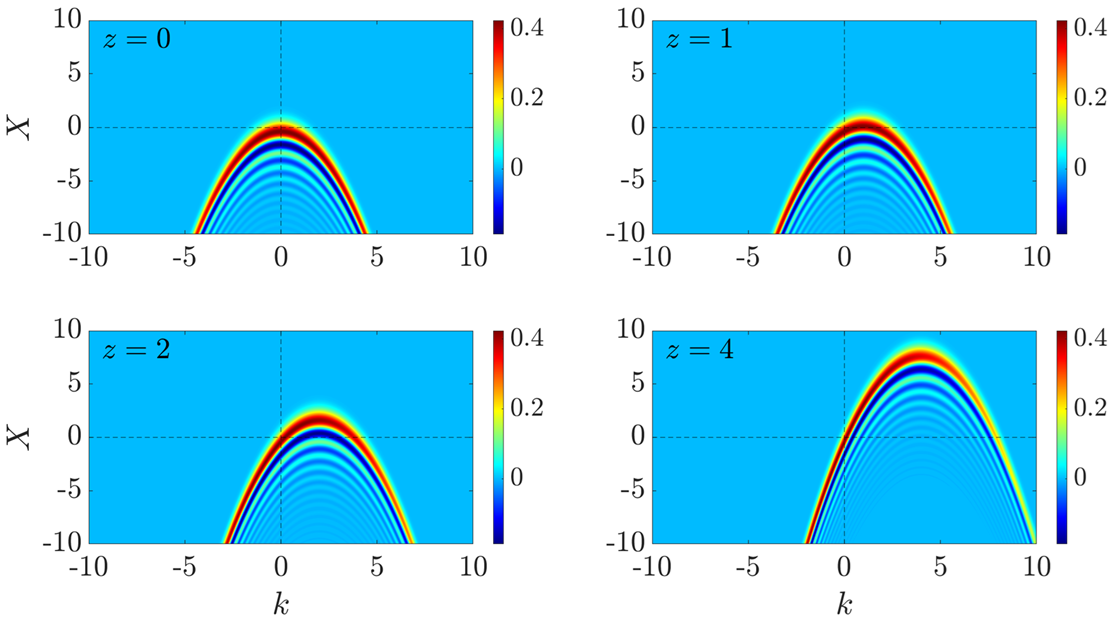}
\caption{Evolution of the WDF of an accelerating wave packet with propagation distance $z$ produced by a fairly coherent source with $\xi_c = 4$. The cut-off parameter of a Gaussian cut-off is $a = 0.1$.}
\label{Fig1}
\end{figure}

\begin{figure}[ht]
\centering
\includegraphics[width=\linewidth]{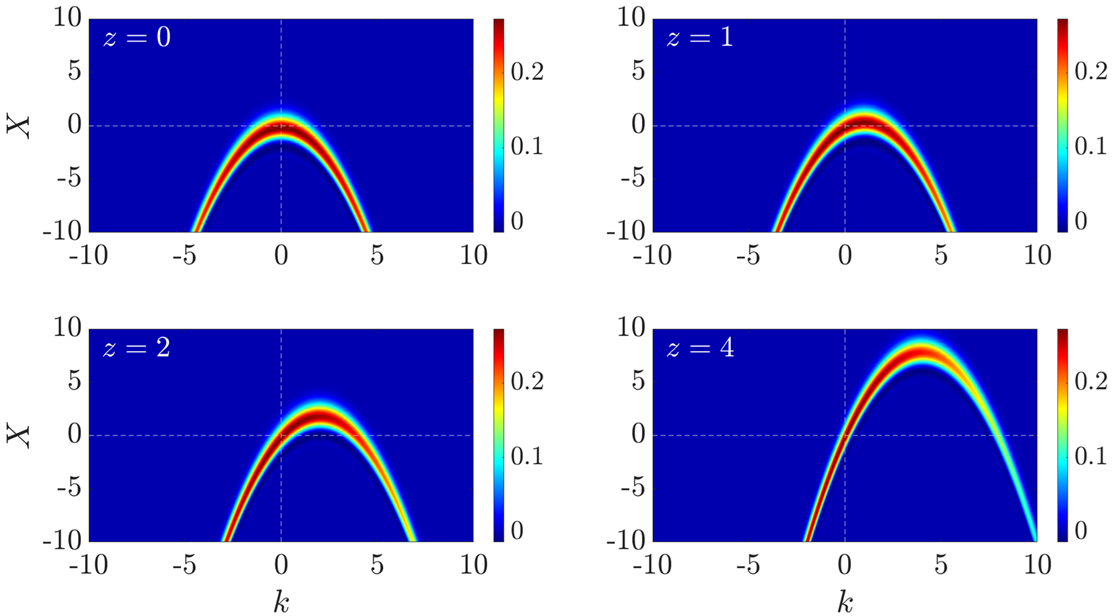}
\caption{Evolution of the  WDF of an accelerating wave packet with propagation distance $z$ engendered by a relatively incoherent source with $\xi_c = 1$. We employ a Gaussian cut-off with the cut-off parameter $a = 0.1$.}
\label{Fig2}
\end{figure}

So far our analysis and results have been generic to accelerating wave packets of any shape and degree of spatial coherence. Let us now specify to a source power spectrum of the form
\begin{equation}\label{p}
    p(\varepsilon)=\sqrt{\xi_c/\pi}\,e^{-\xi_c\varepsilon^2},
\end{equation}
where $\xi_c$ is a coherence parameter. It follows at once from Eq.~(\ref{p}) that $\xi_c \rightarrow\infty$ corresponds to a coherent wave packet with $p(\varepsilon)=\delta (\varepsilon)$, which yields the standard Wigner function of a coherent Airy beam, first derived in~\cite{chen2011wigner} and corrected in~\cite{besieris2011wigner}. Any finite $\xi_c$ controls the dynamics of this family of partially coherent accelerating beams. Combining Eqs.~(\ref{Wig-gen}), ~(\ref{Wig-fp}), and~(\ref{p}) and performing the integration, we obtain, after some steps spelled out in the Supplement, the following closed-form expression for the Wigner function:
\begin{equation}\label{Wig-exp}
\begin{split}
    \mtW_{\mathrm{G}} (X,k,z)&\propto f_a(k) e^{[2X-z^2+(k-z)^2]/\xi_c}e^{2/3\xi_c^3}
        \\
    &\times Ai[2X-z^2+1/\xi_c^2+(k-z)^2].
\end{split}
\end{equation}
Here the subscript "G" corresponds to a Gaussian power spectrum of the source. We can infer from Eq.~(\ref{Wig-exp}) that the Wigner distribution reveals acceleration in physical space, combined with drift in momentum space. To illustrate this point, we visualize in Fig.~\ref{Fig1} the evolution of the wave packet generated by a fairly coherent source with $\xi_c = 4$ with a Gaussian cut-off function $f_a(k)=e^{-a^2k^2}$ with $a=0.1$. We observe in the figure that for a relatively small cut-off scale $a$, the Wigner distribution remains largely unchanged over the range of propagation distances considered here, indicating robustness against finite-energy truncation during propagation. A qualitatively similar scenario is realized for the WDF produced by a relatively incoherent source with $\xi_c=1$ and the same cut-off which we exhibit in Fig.~\ref{Fig2}. We can conclude comparing Figs.~\ref{Fig1} and~\ref{Fig2} that the peak of the WDF is reduced as the source coherence decreases. This conclusion is supported by Eq.~(\ref{Wig-exp}) that features an exponential cut-off to the WDF amplitude due to reduced coherence of the source.  In addition, finite coherence provides a natural exponential cut-off point for the WDF oscillations: The domain of negative values of $\mtW$, which herald coherent interference effects, shrinks with $\xi_c$. Despite the absence of oscillatory features in the low-coherence source case, though, the
WDF  preserves its overall profile as it propagates
away from the source, at least, over four characteristic diffraction lengths. We also studied several cut-off functions and discovered that the WDF evolution is virtually insensitive to the shape of the momentum cut-off so long as the cut-off parameter is sufficiently small.\\

\begin{figure}[ht]
\centering
\includegraphics[width=\linewidth]{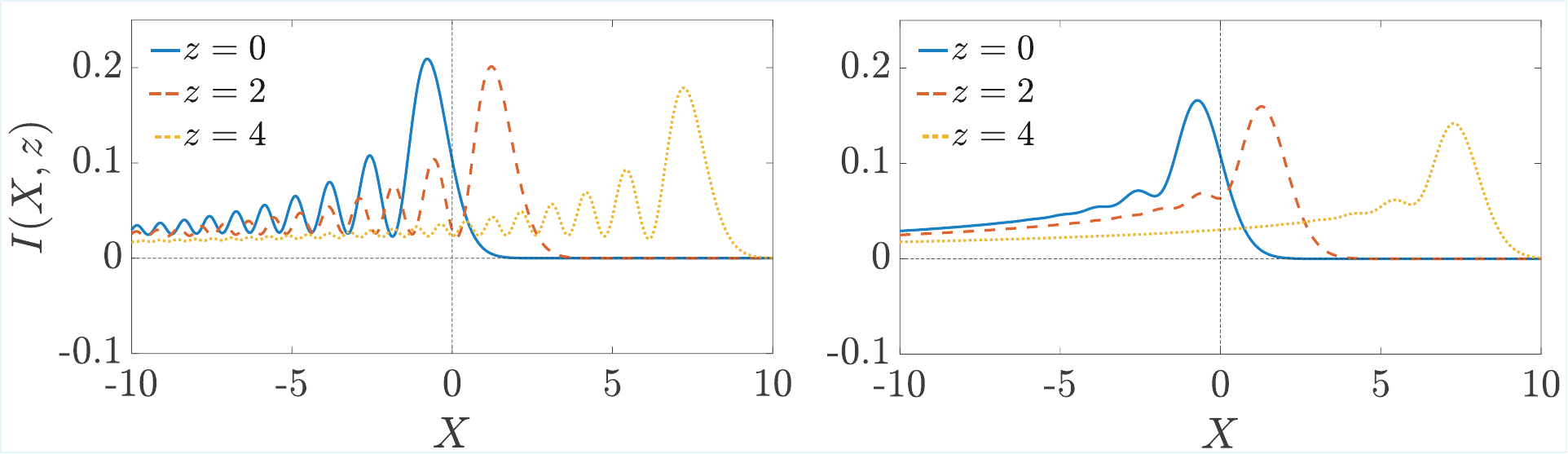}
\caption{Intensity profiles of accelerating wave packets generated by highly coherent $\xi_c = 10$ (left) and moderately incoherent $\xi_c = 2$ (right) sources as functions of X for three propagation distances: $z = 0$ (solid curve), $z = 2$ (dashed curve), and $z = 4$ (dotted curve). The cut-off parameter of a Gaussian cut-off is $a = 0.1$.}
\label{Fig3}
\end{figure}

Further, the intensity profile of any accelerating random Airy wave packet can be numerically evaluated using Eqs.~(\ref{I-def}) and~(\ref{Wig-exp}). In particular, for the same Gaussian cut-off with the parameter $a=0.1$, we display said intensity profile in Fig.~\ref{Fig3}. We can infer from the figure that a highly coherent source with $\xi_c = 10$ yields pronounced oscillations in the left tail of the intensity profile, while much less coherent source with $\xi_c = 2$ gives rise to a nearly monotonic intensity profile. As each wave packet propagates along the $z$-axis, the power is redistributed between the peak of the main lobe and the left oscillatory or monotonous tail of the packet. We also notice that regardless of the state of coherence of the source, the overall shape of the accelerating wave packet remains virtually intact on propagation over, at least, a few characteristic diffraction lengths associated with a typical transverse scale of the source, such as the width of the main lobe, say. 

In conclusion, we have advanced a phase-space theory of generic accelerating, non-diffracting paraxial wave packets of arbitrary spatial coherence using the Wigner distribution function. The WDF evolution of any such wave packet exhibits acceleration in physical space and uniform drift in momentum space. We have also demonstrated how the WDF of any accelerating wave packet of finite-power can be constructed. The theory furnishes an elegant, closed-form representation of the WDF of a family of finite-power accelerating wave packets with a Gaussian energy spectrum of noise.  For any finite source coherence, the WDF displays robustness against finite-energy truncation, preserving its overall phase-space structure during propagation whenever a characteristic momentum space cut-off is sufficiently small. Our results may enable applications to particle steering and guiding, as well as to optical communications through noisy and randomly varying environments. In addition, there has been notable recent progress in optical image encoding and transfer through very noisy environments, including the turbulent atmosphere, by leveraging spatial coherence of light sources; see, for example,~\cite{shen2021optical,xu2022structurally,li2024deep,liu2025unlocking}. The developed WDF formalism, which treats coherent and partially coherent fields in a unified fashion, can facilitate further advances in this direction.

\begin{backmatter}
\bmsection{Funding} 
Natural Sciences \& Engineering Research Council (NSERC) of Canada (RGPIN-2025-04064).

\bmsection{Acknowledgments} 
SAP acknowledges support from NSERC (RGPIN-2025-04064). MH acknowledges support through an Izaak Walton Killam Predoctoral Fellowship.

\bmsection{Disclosures} 
The authors declare no conflicts of interest.

\bmsection{Supplemental Document}
See supplemental document for supporting content. 
\end{backmatter}

\bibliography{references}

\bibliographyfullrefs{references}

\newpage

\end{document}